\documentstyle[12pt]{article}
\newcommand{\address}[1]{\centerline{\small\it #1}}
\renewcommand{\title}[1]{\begin{center}%
{\large\bf #1}\end{center}\par\bigskip}
\renewcommand{\author}[1]{\centerline{\large #1}}
\input epsf
\begin{document}
\title{Ferromagnetism in Hubbard Models\footnote{
Phys. Rev. Lett. {\bf 75}, 4678--4681, 1995.
Archived as cond-mat/9509063.
}}
\author{Hal Tasaki$^*$}
\address{Department of Physics, Gakushuin University,
Mejiro, Toshima-ku, Tokyo 171, JAPAN}
\begin{abstract}
We present the first rigorous examples of non-singular Hubbard models 
which exhibit ferromagnetism at zero temperature. 
The models are defined in arbitrary dimensions, and are characterized by 
finite-ranged hoppings, dispersive bands, and finite on-site Coulomb 
interaction $U$.
The picture, which goes back to Heisenberg, that sufficiently large 
Coulomb interaction can revert Pauli 
paramagnetism into ferromagnetism has finally been
confirmed in concrete examples. 
\end{abstract}
\paragraph*{Introductions:}
The origin of ferromagnetism has been a mystery in physical science for 
quite a long time \cite{Mattis81}.
It was Heisenberg \cite{Heisenberg28}
who first realized that ferromagnetism is 
intrinsically a 
quantum many-body effect, and proposed a scenario that 
spin-independent Coulomb interaction and the Pauli exclusion principle 
generate ``exchange interaction'' between electronic spins.
One of the motivations to study the so-called Hubbard model
has been to 
establish and understand the generation of ferromagnetism in simplified 
situations \cite{HubbRev,MullerHartmannNote}. 
Unfortunately, rigorous examples of ferromagnetism 
(or ferrimagnetism) in the Hubbard models 
have been limited to singular models which have infinitely large 
Coulomb interaction (Nagaoka-Thouless ferromagnetism
\cite{Nagaoka}), or in which
magnetization is supported by a dispersionless band
(Lieb's ferrimagnetism \cite{Lieb89}, and 
flat-band ferromagnetism due to Mielke 
\cite{Mielke}
and the present author
\cite{flat}).
In \cite{94c,95b}, local stability of ferromagnetism in a generic family of
Hubbard models with nearly-flat bands was proved.

In the present Letter, we treat a class of Hubbard models
in arbitrary dimensions, which are non-singular in the sense that they 
have finite ranged hoppings,  
dispersive (single-electron) bands, and finite Coulomb interaction $U$.
We prove that the models exhibit ferromagnetism in their ground states 
provided that $U$ is sufficiently large.
We recall that Hubbard models with dispersive bands (like ours)
exhibit Pauli paramagnetism when 
$U=0$, and remain non-ferromagnetic for sufficiently small $U$.
The appearance of ferromagnetism is a purely non-perturbative 
phenomenon.

As far as we know, this is the first time that the existence of 
ferromagnetism is established 
in non-singular itinerant electron systems.
We stress that our examples finally provide the 
definite affirmative answer 
to the long standing
fundamental problem; 
{\em whether spin-independent Coulomb interaction 
can be the origin of ferromagnetism in itinerant electron systems}
\cite{origin}.
See \cite{flat,95b,next} for further discussions on ferromagnetism in the 
Hubbard models.

\paragraph*{Main results:}
In order to simplify the discussion, we describe our results in 
one-dimensional models.
We discuss models in higher dimensions at the end of the Letter.
Let $N$ be an arbitrary integer, and let $\Lambda$ be the set of integers 
$x$ with $|x|\le N$.  We identify $x=-N$ and $x=N$ to regard $\Lambda$ as 
a periodic chain with $2N$ sites.
We denote by ${\cal E}$ and ${\cal O}$ the subsets of $\Lambda$ 
consisting of even and odd sites (integers), respectively.
As usual we denote by $c^\dagger_{x,\sigma}$, $c_{x,\sigma}$, and 
$n_{x,\sigma}$ the creation, the annihilation, and the number operators, 
respectively, for an electron at site $x\in\Lambda$ with spin 
$\sigma=\uparrow,\downarrow$. 

We consider the standard Hubbard Hamiltonian
\begin{equation}
	H=
	\mathop{\sum_{x,y\in\Lambda}}_{\sigma=\uparrow,\downarrow}
	t_{x,y} \, c^\dagger_{x,\sigma}c_{y,\sigma}
	+
	U\sum_{x\in\Lambda}n_{x,\uparrow}n_{x,\downarrow},
	\label{Ham}
\end{equation}
where 
$t_{x,x+1}=t_{x+1,x}=t'$ for any $x\in\Lambda$,
$t_{x,x+2}=t_{x+2,x}=t$ if $x\in{\cal E}$,
$t_{x,x+2}=t_{x+2,x}=-s$ if $x\in{\cal O}$,
and
$t_{x,x}=V$ if $x\in{\cal O}$.
The remaining elements of $t_{x,y}$ are vanishing.
See Figure~1.
Here $s$, $t$, and $U$ are positive parameters \cite{UNote}.
The parameters $t'$ and $V$ are determined by 
$s$, $t$, and another 
positive parameter $\lambda$ as
$t'=\lambda(s+t)$ and  $V=(\lambda^2-2)(s+t)$.
Our main theorem applies to the case $\lambda=\sqrt{2}$, where we have 
$V=0$. 
We consider the Hilbert space with $N$ electrons in the system.
This corresponds to the quarter-filling of the whole bands, or the 
half-filling of the lower band.

\begin{figure}
\epsfxsize=12cm
\centerline{\epsfbox{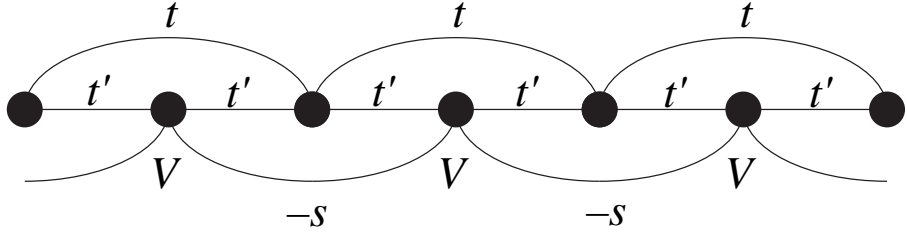}}
\caption[dummy]{
The one-dimensional lattice and the hopping matrix elements.
There are hoppings to nearest and next-nearest neighbors, 
and on-site potential. 
}
\end{figure}

If we consider the single-electron problem corresponding to the 
Hamiltonian (\ref{Ham}), we find that the model has two bands with  
dispersion relations 
$\varepsilon_1(k)=-2s\cos 2k-2(s+t)$, and
$\varepsilon_2(k)=2t\cos 2k+\lambda^2(s+t)$ with $|k|\le\pi/2$.
Note that both the bands have perfect cosine dispersions, which is a 
special feature of the present model \cite{IdealNote}.
There is an energy gap $\lambda^2(s+t)$ between the two bands.

For $\alpha=1,2,3$, we define the total spin operators by 
$S^{(\alpha)}_{\rm tot} = 
\sum_{x\in\Lambda}\sum_{\sigma,\tau=\uparrow,\downarrow}
c^\dagger_{x,\sigma}\,(p^{(\alpha)}/2)_{\sigma,\tau}\,c_{x,\tau}$,
where $p^{(\alpha)}$ are the Pauli matrices, and denote the eigenvalues 
of 
$({\bf S}_{\rm tot})^2=\sum_{\alpha=1}^3(S^{(\alpha)}_{\rm tot})^2$
as $S_{\rm tot}(S_{\rm tot}+1)$.
The maximum possible value of $S_{\rm tot}$ is $S_{\rm max}\equiv N/2$.

Let $b^\dagger_{k,\sigma}$ be the creation operator corresponding to the 
single-electron eigenstate with the energy $\varepsilon_1(k)$.
Let $\Phi_{\rm vac}$ be the state with no electrons.
The state 
$\Phi_{\rm ferro}
=(\prod_k b^\dagger_{k,\uparrow})\Phi_{\rm vac}$
(where the lower band is fully filled by up-spin electrons)
has the lowest energy among the states with $S_{\rm tot}=S_{\rm max}$.
It is easy to observe that $\Phi_{\rm ferro}$ is an eigenstate of $H$ with 
the energy $E_0=-2(s+t)N$.
A simple variational calculation shows that $\Phi_{\rm ferro}$ cannot be a 
ground state of $H$ (and hence the true ground state has  
$S_{\rm tot}<S_{\rm max}$)
if $U<4s$.
The main result of the present Letter is the following.

{\em Theorem I}---Suppose 
$\lambda>\lambda_{\rm c}=[(2+\sqrt{5})^{1/2}-2]^{1/2}\simeq0.241$.
If $t/s$ and $U/s$ are sufficiently large, the ground states of the 
Hamiltonian (\ref{Ham}) have $S_{\rm tot}=S_{\rm max}$, and are 
non-degenerate apart from the $(2S_{\rm tot}+1)$-fold spin degeneracy.
$\Phi_{\rm ferro}$ is one of the ground states.
How large the parameters should be can be determined by diagonalizing 
a Hubbard model on a five-site chain.
(See Figure~2.)

\begin{figure}
\epsfxsize=8cm
\centerline{\epsfbox{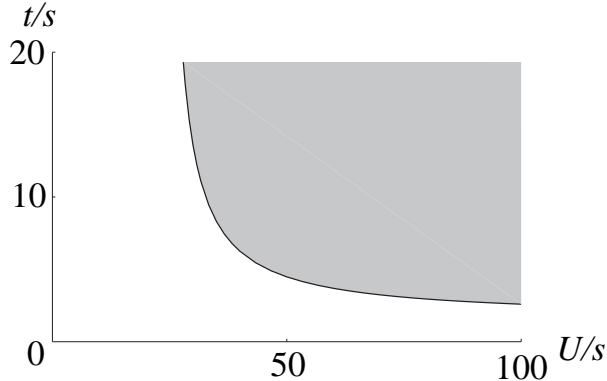}}
\caption[dummy]{
When $\lambda=\sqrt{2}$, Theorem I is applicable for $t/s$ and $U/s$ in 
the shaded region.
The existence of ferromagnetism is established, for examples, if 
$t\ge4.5\,s$ when $U=50\,s$, or $t\ge2.6\,s$ when $U=100\,s$.
Though the plot was obtained from a numerical calculation in a five-site 
Hubbard model, our theorem guarantees that the model with
{\em arbitrary\/} lattice size exhibits ferromagnetism.
}
\end{figure}

The present models reduce to the flat-band models 
studied in \cite{flat} if we set $s=0$.
Therefore we can regard Theorem~I as a confirmation in special cases
of the previous 
conjecture \cite{KusakabeAoki,flat,94c,95b} 
that the flat-band ferromagnetism is  
stable under perturbations.
Moreover the  
strong result about the spin-wave excitation proved in 
\cite{94c,95b} also applies to the preset models.

{\em Theorem II}---Suppose that the model parameters satisfy
$\lambda\ge\lambda_3$,
$s/t\le\rho_0$, and
$K_2\lambda t\ge U\ge A_3\lambda^2 s$,
where $\lambda_3$, $\rho_0$, $K_2$, and $A_3$ are positive constants 
that appear in \cite{95b}. 
Then the spin-wave excitation energy $E_{\rm SW}(k)$ 
(i.e., the lowest energy among the states with $(N-1)$ up-spin electrons 
and one down-spin electron and with crystal momentum $k$)
of the Hubbard model (\ref{Ham}) satisfies 
\begin{equation}
	F_2\frac{4U}{\lambda^4}(\sin k)^2
	\le
	E_{\rm SW}(k)-E_0
	\le
	F_1\frac{4U}{\lambda^4}(\sin k)^2,
	\label{SW}
\end{equation}
where $E_0$ is the ground state energy, and $F_1$, $F_2$ are constants 
such that $F_1\simeq F_2\simeq1$ if $\lambda\gg1$, $\lambda s\ll t$,
and $U\gg\lambda^2 s$.
(See \cite{95b} for details and a proof.)

In the parameter region where both Theorems~I and II are applicable, we 
have an ideal situation that the global stability of ferromagnetism as
well as 
the appearance of ``healthy'' low-lying excited states are  
established rigorously. 
We have rigorously derived a ferromagnetic system with (effective) 
exchange interaction $J\simeq 2U/\lambda^4$, starting from 
the Hubbard 
models for itinerant electrons!

It is quite likely that the present models represent (Mott-Hubbard)
insulators.
We expect that the same models with smaller electron numbers describe 
metallic ferromagnetism \cite{metal}, 
but have no rigorous results in this direction
(except for those in the flat-band models \cite{flat}).

\paragraph*{Proof of Theorem I:}
For $x\in\Lambda$ and $\sigma=\uparrow,\downarrow$, we define
$a_{x,\sigma}=\lambda\,c_{x,\sigma}-
(-1)^{x}(c_{x-1,\sigma}+c_{x+1,\sigma})$, 
which correspond to the strictly localized basis states used in 
\cite{flat,94c,95b}.
The anti-commutator $\{a^\dagger_{x,\sigma},a_{y,\sigma}\}$ is
$\lambda^2+2$ if $x=y$, is $1$ 
if $|x-y|=2$, and is vanishing otherwise.
By using these operators, Hamiltonian (\ref{Ham}) can be written in a 
compact manner as
\begin{equation}
	H=
	(\lambda^2 s -2t)N
	-s\mathop{\sum_{x\in{\cal E}}}_{\sigma=\uparrow,\downarrow}
	a^\dagger_{x,\sigma}a_{x,\sigma}
	+t\mathop{\sum_{x\in{\cal O}}}_{\sigma=\uparrow,\downarrow}
	a^\dagger_{x,\sigma}a_{x,\sigma}
	+U\sum_{x\in\Lambda}n_{x,\uparrow}n_{x,\downarrow},
	\label{Ham2}
\end{equation}
in the sector with $N$ electrons.
We further rewrite it as 
$H=(\lambda^2 s -2t)N+\sum_{x\in{\cal E}}h_x$ with the local 
Hamiltonian defined as
\begin{equation}
	h_x=-s\sum_{\sigma=\uparrow,\downarrow}
	a^\dagger_{x,\sigma}a_{x,\sigma}+U\,n_{x,\uparrow}n_{x,\downarrow}
	+\sum_{r=\pm1}
	\left(
	\frac{t}{2}\sum_{\sigma=\uparrow,\downarrow}
	a^\dagger_{x+r,\sigma}a_{x+r,\sigma}
	+\frac{U}{2}\,n_{x+r,\uparrow}n_{x+r,\downarrow}
	\right).
	\label{hx}
\end{equation}
Since $[h_x,h_{x+2}]\ne0$, it is impossible to diagonalize all $h_x$ 
simultaneously.

{\em Lemma}---Suppose  
$\lambda>\lambda_{\rm c}$, and 
$t/s$ and $U/s$ are sufficiently large.
Then the minimum eigenvalue of $h_x$ 
(regarded as an operator on the whole Hilbert space) is
$-(\lambda^2+2)s$.
In any of the corresponding eigenstates,
there are one, two, or three electrons in the sublattice 
$\{x-2,x-1,x,x+1,x+2\}$, and these electrons are coupled 
{\em ferromagnetically}.
Any eigenstate $\Phi$ with the eigenvalue $-(\lambda^2+2)s$ can 
be written in the form 
\begin{equation}
	\Phi=
	a^\dagger_{x,\uparrow}\Phi_1
	+a^\dagger_{x,\downarrow}\Phi_2,
	\label{Phicond1}
\end{equation}
with some states $\Phi_1$ and $\Phi_2$, and satisfies
\begin{equation}
	c_{x+1,\downarrow}c_{x+1,\uparrow}\Phi
	=0.
	\label{Phicond2}
\end{equation}

We prove the Lemma in the next part.
In what follows, we assume that the model parameters satisfy the 
conditions 
in the Lemma.
The basic strategy of the proof of Theorem I 
is to extend the local ferromagnetism 
found above into a global ferromagnetism.
Special characters of the present model makes such an extension 
possible.

The Lemma implies $h_x\ge-(\lambda^2+2)s$, and hence 
$H\ge(\lambda^2 s -2t)N-N(\lambda^2+2)s=
-2(s+t)N=E_0$.
This proves that $\Phi_{\rm ferro}$ 
(which has the eigen-energy $E_0$) is a ground state.

To show the uniqueness of the ground states, we assume $\Phi$ is a ground 
state, i.e., $H\Phi=E_{0}\Phi$.
Then we  have $h_x\Phi=-(\lambda^2+2)s\,\Phi$ for each $x\in{\cal E}$,
and $\Phi$ is characterized by the Lemma.
We note that the collection of states 
$(\prod_{x\in A}a^\dagger_{x,\uparrow})
(\prod_{x\in B}a^\dagger_{x,\downarrow})\Phi_{\rm vac}$
with arbitrary subsets $A,B\subset\Lambda$ such that 
$|A|+|B|=N$ forms a (complete) basis of the $N$-electron Hilbert space.
Imagine that we expand $\Phi$ using this basis.
Since (\ref{Phicond1}) holds for any $x\in{\cal E}$,  
$\Phi$ must be written in the form
\begin{equation}
	\Phi=\sum_{\tilde{\sigma}}\varphi(\tilde{\sigma})
	(\prod_{x\in{\cal E}}a^\dagger_{x,\sigma(x)})
	\Phi_{\rm vac},
	\label{Phi=}
\end{equation}
where $\tilde{\sigma}=(\sigma(x))_{x\in{\cal E}}$ 
is a spin configuration with 
$\sigma(x)=\uparrow,\downarrow$, and $\varphi(\tilde{\sigma})$ 
is a coefficient. 
Unlike in the flat-band models \cite{flat}, 
a state of the form (\ref{Phi=}) is not 
necessarily an eigenstate of the hopping part of $H$.

By examining how $c_{x+1,\downarrow}c_{x+1,\uparrow}$ acts on (\ref{Phi=}),
the condition (\ref{Phicond2}) reduces to
\begin{equation}
	\mbox{$\varphi(\tilde{\sigma})=\varphi(\tilde{\sigma}_{x,x+2})$
   for any $\tilde{\sigma}$,}
	\label{symmetric}
\end{equation}
where $\tilde{\sigma}_{x,x+2}$ is the spin configuration 
obtained by switching $\sigma(x)$ and $\sigma(x+2)$ in $\tilde{\sigma}$.
Since (\ref{symmetric}) holds for any $x\in{\cal E}$, we find that 
$\varphi(\tilde{\sigma})=\varphi(\tilde{\tau})$ whenever 
$\sum_{x\in{\cal E}}\sigma(x)=\sum_{x\in{\cal E}}\tau(x)$.
Since  $\Phi_{\rm ferro}$ is written as
$\Phi_{\rm ferro}=
{\rm const.} (\prod_{x\in{\cal E}}a^\dagger_{x,\uparrow})
\Phi_{\rm vac}$,
this means that $\Phi$ can be written in the form
$\Phi=\sum_{M=0}^N\alpha_M(S^-_{\rm tot})^M\Phi_{\rm ferro}$,
where the spin lowering operator is 
$S^-_{\rm tot}=\sum_{x\in\Lambda}c^\dagger_{x,\downarrow}c_{x,\uparrow}$.
This proves that $\Phi_{\rm ferro}$ and its $SU(2)$ rotations are the only 
ground states of $H$.

\paragraph*{Proof of Lemma:}
Because of the translation invariance, it suffices to prove
the Lemma for $x=0$.
We first diagonalize the hopping part of $h_0$ (obtained by 
setting $U=0$).
We express a single-electron state supported on the sublattice
$\Lambda_0=\{-2,-1,0,1,2\}$ as a 
five-dimensional vector 
$\varphi=(\varphi_{-2},\varphi_{-1},\varphi_{0},\varphi_{1},
\varphi_{2})$.
The normalized eigenstates are
$\varphi^{(0)}=(\lambda^2+2)^{-1/2}(0,-1,\lambda,-1,0)$
with the eigenvalue $\varepsilon_0=-(\lambda^2+2)s$,
$\varphi^{(1)}=\{2(\lambda^2+1)\}^{-1/2}(\lambda,-1,0,1,-\lambda)$
with $\varepsilon_1=0$,
$\varphi^{(2)}=\{2(\lambda^2+2)(\lambda^2+3)\}^{-1/2}
(-(\lambda^2+2),\lambda,2,\lambda,-(\lambda^2+2))$
with $\varepsilon_2=0$,
and two more with $\varepsilon_3=(\lambda^2+1)t/2$ and
$\varepsilon_4=(\lambda^2+3)t/2$.
We denote the corresponding creation operators as
$d^\dagger_{i,\sigma}=
\sum_{x\in\Lambda_0}\varphi^{(i)}_xc^\dagger_{x,\sigma}$.
It is crucial to note that 
$a^\dagger_{0,\sigma}=(\lambda^2+2)^{1/2}d^\dagger_{0,\sigma}$.

Since the local Hamiltonian $h_0$ conserves the number of electrons 
in $\Lambda_0$, we can examine its minimum 
eigenvalue in each sector with a fixed number of electrons in $\Lambda_0$. 
When there are no electrons in $\Lambda_0$, the only possible eigenvalue 
of $h_0$ is $0>-(\lambda^2+2)s$.
Let $f_n$ and $e_n$ be the minimum eigenvalues of $h_0$ in the sectors 
with 
$n$ electrons in $\Lambda_0$ with the total spin (of the $n$ electrons) 
$S^{(n)}_{\rm tot}=n/2$, and $S^{(n)}_{\rm tot}<n/2$, respectively.

Noting that
$f_n=\sum_{m=0}^{n-1}\varepsilon_m$, we find 
$f_1=f_2=f_3=-(\lambda^2+2)s$ and
$f_n>-(\lambda^2+2)s$ for $n\ge4$.
Since the corresponding ferromagnetic eigenstates are
$(\prod_{m=0}^{n-1}d^{\dagger}_{m,\uparrow})\widetilde{\Phi}$
(where $\widetilde{\Phi}$ is an arbitrary state with no electrons in 
$\Lambda_{0}$) or their $SU(2)$ rotations, they are
written in the desired form (\ref{Phicond1}), 
and satisfy (\ref{Phicond2}).
Therefore, in order to prove the Lemma, it suffices 
to show  
\begin{equation}
	\mbox{$e_n > -(\lambda^2+2)s$ for any $n=2,3,\ldots,8$.}
	\label{thecond}
\end{equation}

Since the condition (\ref{thecond}) only involves eigenvalues of a finite 
system, it can be checked by numerically diagonalizing finite 
dimensional matrices for given values of $\lambda$, $s$, $t$ and $U$.
We can thus construct  a {\em computer aided 
proof\/} that our Hubbard model exhibits ferromagnetism.
Figure~2 summarizes the result of a preliminary analysis in this direction.

Let us prove (\ref{thecond}) in a range of parameters
without using computers.
Let $e_2^{\rm sym}$ ({\em resp.}, $e_2^{\rm as}$) be
the minimum eigenvalue of $h_0$ in the sector with two electrons in 
$\Lambda_0$ forming spin-singlet states which is symmetric
({\em resp.}, antisymmetric) under the spatial
reflection $x\rightarrow-x$.
Let us evaluate  $e_2^{\rm sym}$.
In the limit $t\uparrow\infty$, a spin-singlet state with two 
electrons in the symmetric sector which has 
finite expectation value of $h_0$ is written as 
\begin{equation}
	\Psi=\left\{
	\alpha\,d^\dagger_{0,\uparrow}d^\dagger_{0,\downarrow}
	+\frac{\beta}{\sqrt{2}}
	(d^\dagger_{0,\uparrow}d^\dagger_{2,\downarrow}
	+d^\dagger_{2,\uparrow}d^\dagger_{0,\downarrow})
	+\gamma\,d^\dagger_{2,\uparrow}d^\dagger_{2,\downarrow}
	+\delta\,d^\dagger_{1,\uparrow}d^\dagger_{1,\downarrow} 
	\right\}
	\widetilde{\Phi},
\end{equation}
where $\widetilde{\Phi}$ is any state with no electrons in $\Lambda_0$.
The expectation value of the hopping part of $h_{0}$ in this state 
is given by
$\langle h_{0}^{\rm hop}\rangle
	=
	(\Psi,h_{0}^{\rm hop}\,\Psi)/(\Psi,\Psi)
	=
	A [-2(\lambda^2+2)s|\alpha|^2
	-(\lambda^2+2)s|\beta|^2]$,
where 
$A=(|\alpha|^{2}+|\beta|^{2}+|\gamma|^{2}+|\delta|^{2})^{-1}$.
If we further let $U\uparrow\infty$, a finite energy state must 
also satisfy 
$c_{0,\downarrow}c_{0,\uparrow}\Psi=0$ and
$c_{1,\downarrow}c_{1,\uparrow}\Psi=0$.
These conditions lead us to the constraints
\begin{eqnarray}
	&&
	\frac{\lambda^2}{\lambda^2+2}\alpha
	+\frac{2\lambda}{(\lambda^2+2)\sqrt{\lambda^2+3}}\beta
	+\frac{2}{(\lambda^2+2)(\lambda^2+3)}\gamma
	=0,
	\nonumber\\
	&&
	\frac{1}{\lambda^2+2}\alpha
	-\frac{\lambda}{(\lambda^2+2)\sqrt{\lambda^2+3}}\beta
	+\frac{\lambda^2}{2(\lambda^2+2)(\lambda^2+3)}\gamma
	+\frac{1}{2(\lambda^2+1)}\delta
	=0.
	\label{abcd}
\end{eqnarray}
We denote the desired minimum 
eigenvalue $e_2^{\rm sym}$ for $t=U=\infty$ as 
$e_{2,\infty}^{\rm sym}$.
To get $e_{2,\infty}^{\rm sym}$, we
minimize the energy expectation value $\langle h_{0}^{\rm hop}\rangle$
with respect to the constraints (\ref{abcd}).
The rest is a tedious but straightforward estimate.
Eliminating $\beta$ from (\ref{abcd}), we get
$\alpha+\gamma/(\lambda^{2}+3)+\delta/(\lambda^{2}+1)=0$,
which implies
$|\gamma|^{2}+|\delta|^{2}\ge f(\lambda)|\alpha|^{2}$
with
$f(\lambda)=\{(\lambda^{2}+1)^{-2}+(\lambda^{2}+3)^{-2}\}^{-1}$.
By substituting this bound into
$\langle h_{0}\rangle+(\lambda^{2}+2)s=
A(\lambda^{2}+2)s(-|\alpha|^{2}+|\gamma|^{2}+|\delta|^{2})$,
we get
$e_{2,\infty}^{\rm sym}+(\lambda^{2}+2)s
\ge 
A(\lambda^{2}+2)s|\alpha|^{2}\{f(\lambda)-1\}$.
Noting that $\alpha\ne0$ in the minimizer, the condition
$f(\lambda)>1$ (which is equivalent to $\lambda>\lambda_{\rm c}$)
implies $e_{2,\infty}^{\rm sym}>-(\lambda^2+2)s$.
Since $e_2^{\rm sym}$ is a continuous function of $t$ and
$U$, this  proves that $e_2^{\rm sym}>-(\lambda^2+2)s$ for 
sufficiently large $t$ and $U$.

By repeating the similar (but easier) variational analysis, we find that 
$e_2^{\rm as}=-(\lambda^2/3)s>-(\lambda^2+2)s$, and
$e_n = \infty$ for $n\ge3$ when $t=U=\infty$.
This implies that the desired condition (\ref{thecond}) holds for 
$\lambda>\lambda_{\rm c}$ and sufficiently large $t$ and $U$.
The Lemma has been proved.

\paragraph*{Models in higher dimensions:}
Models in higher dimensions can be constructed and analyzed in quite the 
same spirit \cite{next}. 
Take, for example, the flat-band models studied in \cite{flat}. 
($V$ and $M$ correspond to ${\cal E}$ and ${\cal O}$ of the present 
paper, respectively.)
For $x\in V$ we let $a_{x,\sigma}$ as in \cite{flat}.
For $x=m(v,w)\in M$, we let 
$a_{x,\sigma}=\lambda c_{x,\sigma}+c_{v,\sigma}+c_{w,\sigma}$.
We define the Hamiltonian as
\begin{equation}
	H=
	-s\mathop{\sum_{x\in V}}_{\sigma=\uparrow,\downarrow}
	a^\dagger_{x,\sigma}a_{x,\sigma}
	+t\mathop{\sum_{x\in M}}_{\sigma=\uparrow,\downarrow}
	a^\dagger_{x,\sigma}a_{x,\sigma}
	+U\sum_{x\in\Lambda}n_{x,\uparrow}n_{x,\downarrow},
	\label{Hamd}
\end{equation}
which again contains next-nearest neighbor hoppings and on-site potentials.
(\ref{Hamd}) should be compared with (\ref{Ham2}).
By a straightforward extension of the present method, 
we can prove that the ground states of the model with $|V|$-electrons 
exhibit ferromagnetism when $\lambda$, $t/s$, and $U/s$ are 
sufficiently large \cite{next}.

\par\bigskip
It is a pleasure to thank 
Tohru Koma,
Andreas Mielke,
and
Bruno Nachtergaele
for useful discussions.

\par\bigskip
{\bf Note added (August 1997):}
The proof of Theorem I has been considerably improved.
Now the condition $\lambda>\lambda_{\rm c}$ has been replaced simply 
by $\lambda\ne0$.
Full details will appear in \cite{next}, which is still under preparation for the 
moment.




\begin{thebibliography}{10}
\bibitem{origin}
The reader might ask what the basic mechanism of 
ferromagnetism in the present models is.
Our ferromagnetism may be (partially) interpreted in terms 
either of the standard two pictures of ferromagnetism, namely, the 
Heisenberg-like ``exchange'' picture and the band-electron picture.
The most crucial point in our proof is that we describe electronic 
states in such a way that both the particle-like character of 
electrons and the band structure of the model are taken into account.
\bibitem{UNote}
We can of course consider models in which ${\cal E}$-sites and 
${\cal O}$-sites have different values of $U$.
\bibitem{IdealNote}
This special property of the dispersion relations is closely related to the
representation (\ref{Ham2}) of the Hamiltonian.
\bibitem{metal}
A support to this conjecture is given by a perturbation theory using 
the Wannier basis.
It suggests that the low energy effective theories of these doped models 
are ferromagnetic $t$-$J$ models, where $|t/J|$ can be made 
arbitrarily small by taking Hubbard models close to flat-band models.
\end{thebibliography}
\end{document}